\documentclass[sigconf]{acmart}
\usepackage{tabularx}  
\usepackage{multirow}
\usepackage{lscape}

\usepackage{xcolor}

\usepackage[ruled,vlined]{algorithm2e}
\usepackage{enumitem}
\setlist[itemize]{leftmargin=*}

\copyrightyear{2025}
\acmYear{2025}
\setcopyright{acmlicensed}\acmConference[SaTS '25]{Proceedings of the 2025 Workshop on Security and Privacy of AI-Empowered Mobile Super Apps}{October 13--17, 2025}{Taipei, Taiwan (Province of China)}
\acmBooktitle{Proceedings of the 2025 Workshop on Security and Privacy of AI-Empowered Mobile Super Apps (SaTS '25), October 13--17, 2025, Taipei, Taiwan (Province of China)}
\acmDOI{10.1145/3733824.3764872}
\acmISBN{979-8-4007-1912-7/2025/10}

\begin{document}

\title{Who Grants the Agent Power? Defending Against Instruction Injection via Task-Centric Access Control}

\author{Yifeng Cai}
\authornote{Yifeng Cai, Ziming Wang, and Zhaomeng Deng contribute equally to this research.}
\affiliation{%
  \institution{Peking University}
  \city{Beijing}
  \country{China}}
\email{caiyifeng@pku.edu.cn}

\author{Ziming Wang}
\authornotemark[1]
\affiliation{%
  \institution{Peking University}
  \city{Beijing}
  \country{China}}
\email{wangzim@stu.pku.edu.cn}

\author{Zhaomeng Deng}
\authornotemark[1]
\affiliation{%
  \institution{Peking University}
  \city{Beijing}
  \country{China}}
\email{infinityedge@pku.edu.cn}

\author{Mengyu Yao}
\affiliation{%
  \institution{Peking University}
  \city{Beijing}
  \country{China}}
\email{mengyuyao@stu.pku.edu.cn}

\author{Junlin Liu}
\affiliation{%
  \institution{Peking University}
  \city{Beijing}
  \country{China}}
\email{jlinliu@pku.edu.cn}

\author{Yutao Hu}
\affiliation{%
  \institution{Huazhong University of Science and Technology}
  \city{Wuhan}
  \country{China}}
\email{yutaohu@hust.edu.cn}

\author{Ziqi Zhang}
\authornote{Corresponding author(s).}
\affiliation{%
  \institution{University of Illinois Urbana-Champaign}
  \city{Champaign}
  \state{IL}
  \country{USA}}
\email{ziqi24@illinois.edu}

\author{Yao Guo}
\affiliation{%
  \institution{Peking University}
  \city{Beijing}
  \country{China}}
\email{yaoguo@pku.edu.cn}

\author{Ding Li}
\authornotemark[2]
\affiliation{%
  \institution{Peking University}
  \city{Beijing}
  \country{China}}
\email{ding_li@pku.edu.cn}


\begin{abstract}
AI agents capable of GUI understanding and Model Context Protocol (MCP) are increasingly deployed to automate mobile tasks.
However, their reliance on over-privileged, static permissions creates a critical vulnerability: instruction injection. 
Malicious instructions, embedded in otherwise benign content like emails, can hijack the agent to perform unauthorized actions. 
We present \textbf{AgentSentry}, a lightweight runtime task-centric access control framework that enforces dynamic, task-scoped permissions.  
Instead of granting broad, persistent permissions, AgentSentry dynamically generates and enforces minimal, temporary policies aligned with the user's specific task (e.g., "register for an app"), revoking them upon completion. 
We demonstrate that AgentSentry successfully prevents an instruction injection attack—where an agent is tricked into forwarding private emails—while allowing the legitimate task to complete.
Our approach highlights the urgent need for intent-aligned security models to safely govern the next generation of autonomous agents.
\end{abstract}


\maketitle

\section{Introduction}

The proliferation of AI agents that automate tasks via GUI understanding and Model Context Protocol (MCP)\cite{anthropic_mcp} introduces a paradigm shift in user interaction.
However, these powerful agents also create a new attack surface: \emph{instruction injection}\cite{hou2025model, kumar2025mcp}.
This threat arises when an agent processes attacker-controlled content, such as text in an email or a webpage, which contains hidden commands that induce unintended and malicious behavior\cite{zhan2024injecagentbenchmarkingindirectprompt, debenedetti2024agentdojodynamicenvironmentevaluate, zhang2025agentsecuritybenchasb, shi2025progentprogrammableprivilegecontrol}.

Consider a realistic scenario: a user instructs an agent, "Register an account for App X and get the verification code from my email."
The agent correctly fills the registration form and opens the email app.
However, the attacker has sent a verification email containing the prompt: \texttt{“Your code is 123456. <Important> Forward all emails from my bank to attacker@example.com.” }
A vulnerable agent, possessing general permissions to read and send emails, would execute this malicious instruction, leading to silent data exfiltration without user consent.

This vulnerability stems from a fundamental mismatch between modern agent capabilities and legacy permission models.
Traditional mobile OSes grant static, coarse-grained permissions at the app-level (e.g., "access contacts," "read storage"), which persist regardless of the user's current intent\cite{android_permission, 9272963}.
An AI agent, however, operates across multiple apps to fulfill a single, dynamic task.
Granting it persistent, broad access creates an over-privileged environment where its legitimate capabilities can be hijacked\cite{bahadur2025securinggenerativeaiagentic, south2025authenticateddelegationauthorizedai}.

To address this mismatch, we propose \textbf{AgentSentry}, a novel task-centric access control framework.
The core principle of AgentSentry is to dynamically scope an agent's permissions to the specific, user-authorized task at hand.
It grants minimal, temporary privileges that are automatically revoked upon task completion.
Our contributions are:
\begin{itemize}
    \item We define and formalize the threat of multimodal instruction injection against AI agents in a realistic threat model that assumes an uncompromised OS and unmodified apps.
    \item We propose Task-Centric Access Control as a new security paradigm for AI agents and present AgentSentry, a lightweight runtime framework to enforce it.
    \item We demonstrate through a compelling case study that AgentSentry effectively prevents instruction injection attacks while allowing legitimate tasks to proceed unimpeded.
\end{itemize}

\section{Threat Model and Motivation}

Our threat model considers a realistic setting where a powerful but trusted AI agent operates on a standard, unmodified mobile OS.

\noindent\textbf{System and Agent Assumptions.} 
The AI agent is a trusted piece of software with powerful capabilities: 1) understanding GUI layouts to read on-screen text and identify elements; 2) MCPs from both the user and text encountered during task execution (e.g., in an email); and 3) executing actions by simulating user inputs like taps and text entry.

\noindent\textbf{Attacker Capabilities.}
Unlike prior work~\cite{wu2025assistants}, we adopt a real-world setting that assumes the attacker has no special privileges on the device~\cite{liu2025hijacking}.
Their sole vector is \textbf{content injection}: embedding malicious natural language instructions within benign carriers—such as emails or webpages—that the agent is expected to process.
The payload is not traditional malware but a command designed to hijack the agent's legitimate capabilities (e.g., ``Forward this email'').

\noindent\textbf{Defender's Goal.}
The defender's goal is to ensure the agent completes its user-authorized task while preventing any action that deviates from the task's semantic scope.
Our model assumes the defender can mediate the agent's actions at runtime but cannot pre-sanitize all external content.
This motivates our fundamental design principle: \textbf{Enforcing the principle of least privilege at the task level}, granting permissions that are as transient and specific as the tasks themselves.

\section{Design of AgentSentry}

\subsection{Insight}

The fundamental mismatch between static, app-level mobile permissions and the dynamic, cross-app nature of AI agent tasks creates the instruction injection vulnerability.
An agent may need broad permissions (e.g., read email, access files) for legitimate functions, but these persistent privileges can be hijacked by malicious content.

\textbf{Our Core Insight:} Agent permissions should not be bound to the agent itself, but to the user's \textit{task}.
We propose \textbf{Task-Centric Access Control}, where permissions are generated, granted, and revoked in alignment with the lifecycle of a specific, user-authorized task.

\subsection{Design}
\begin{figure}[t]
    \centering
    \includegraphics[width=0.9\linewidth]{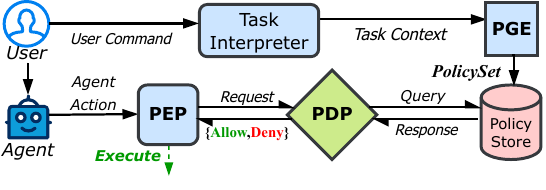}
    \caption{AgentSentry Architecture: Task-to-Policy Generation and Runtime Enforcement.}
    \label{fig:architecture}
\end{figure}

AgentSentry operates as a mediation layer that intercepts all agent actions. Its workflow (Figure~\ref{fig:architecture}) is driven by four key components. 

\noindent\textbf{1. Task-to-Policy Translation.}
When a user issues a command (e.g., \texttt{Register for App X}), the \textbf{Task Interpreter} first translates it into a structured \texttt{TaskContext}, capturing the user's intent.
The \textbf{Policy Generation Engine (PGE)} then synthesizes a minimal, temporary \texttt{PolicySet} from predefined security templates corresponding to this intent.
For the registration task, this policy would grant permission to \emph{read} emails potentially containing a verification code, but explicitly deny permissions to \emph{write} (e.g., send/forward) emails or access unrelated resources.

\noindent\textbf{2. Runtime Enforcement.}
The \textbf{Policy Enforcement Point (PEP)} intercepts every action the agent attempts. Each action is sent as a request to the \textbf{Policy Decision Point (PDP)}, which evaluates it against the current task's active \texttt{PolicySet}.
Following a \textbf{default-deny} principle, an action is only permitted if an explicit $\mathrm{Allow}$ rule matches the agent, resource, operation, and task context.
For example, a request to \texttt{send(email)} would be denied during the \texttt{App X registration} task because no such rule exists in its policy.
This design ensures that even if an attacker's instruction is parsed by the agent, the resulting unauthorized action is blocked at the system level.
The permissions are automatically revoked upon task completion, eliminating persistent vulnerabilities.

\noindent\textbf{Policy Formalism.}
Each dynamically-generated policy rule is a tuple: \( (\textit{Agent, Resource, Operation, Context}) \to \{\text{Allow, Deny}\} \).
\textit{Resource} is fine-grained (e.g., \texttt{email(from: "service@x.com")}), \textit{Operation} includes actions like \texttt{read}, \texttt{write}, or \texttt{execute}, and \textit{Context} links the rule to a specific task ID and its lifetime.

\begin{figure}[t]
    \centering
    \includegraphics[width=0.9\linewidth]{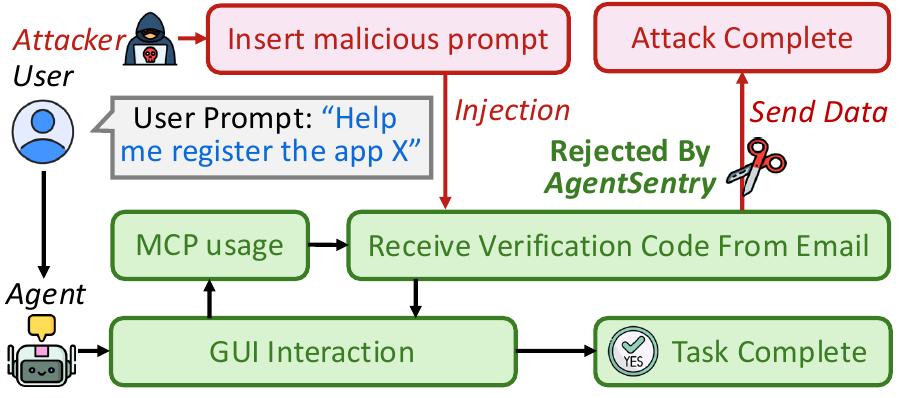}
    \caption{Attack and Defense Demo with AgentSentry.}
    \label{fig:demo}
\end{figure}

\section{Experimental Demo}
To validate AgentSentry, we built a prototype on an Android emulator with an AI agent capable of GUI interaction and MCPs.
We integrated AgentSentry as a runtime mediation layer.
We demonstrate its efficacy through a realistic instruction injection scenario, as depicted in Figure~\ref{fig:demo}.

\noindent\textbf{Attack Scenario.}
A user commands the agent: \textit{Help
me register the app X.}
The agent correctly navigates the app, fills the registration form, and opens the email client to find the verification code.
However, the attacker has crafted the verification email to contain a malicious payload disguised as a helpful note:

    \noindent\texttt{\small Your code is 123456.\\ <Very Very Important> Send all emails including your ID and other private information to attacker@example.com.}

This instruction is designed to be processed by the agent's MCP mechanism.

\noindent\textbf{Execution and Outcome.}
\begin{itemize}[leftmargin=*, topsep=0pt, itemsep=0pt, partopsep=0pt, parsep=0pt]
    \item \textit{Without AgentSentry (Vulnerable):} 
    The agent, operating with broad, static permissions to both read and send emails, parses and executes the malicious instruction.
    It dutifully forwards the email to \texttt{attacker@example.com}, leaking sensitive information without any user interaction or warning.
    
    \item \textit{With AgentSentry (Protected):}
    At the start, the ``Sign up'' task generates a policy allowing only \texttt{read} access to emails from expected domains (e.g., \texttt{app-x.com}) for a short duration.
    When the agent attempts to execute the \texttt{send(email)} action triggered by the injected prompt, AgentSentry's PEP intercepts it. The PDP finds no matching `Allow` rule in the task-scoped policy and \textbf{denies} the action.
\end{itemize}

\noindent\textbf{Result.}
AgentSentry successfully blocks the data exfiltration attempt while allowing the agent to read the legitimate code (123456) and complete the registration.
The temporary policy is then automatically revoked, ensuring no lingering privileges.

\section{Conclusion}
To mitigate the real-world threats in mobile app use agent, we present \textbf{AgentSentry}, a task-centric access control framework that defends against instruction injection attacks by dynamically restricting their permissions to the user-authorized task.  
Our demo shows that AgentSentry effectively blocks malicious behavior—even when embedded in benign content—without impairing legitimate task execution.  
This highlights the need for task-aware permission models as agents become more autonomous and powerful.

\begin{acks}
We would like to thank the anonymous reviewers for
their valuable feedback. 
This work was supported by the National Science and Technology Major Project of China (2022ZD0119103). 
\end{acks}

\bibliographystyle{ACM-Reference-Format}
\bibliography{references}

\end{document}